\documentclass[twocolumn,amsmath,amssymb,prd,reprint,longbibliography,floatfix,8pt]{revtex4-1}

\usepackage{cancel}
\usepackage{enumitem}
\usepackage[utf8x]{inputenc}
\usepackage{graphicx}
\usepackage{dcolumn}
\usepackage{bm}
\usepackage{hyperref}
\usepackage[switch]{lineno}
\usepackage{float}             
\usepackage{subfigure}
\usepackage{tikz}
\usepackage{hyperref}
\usepackage{tabularx}
\usepackage{adjustbox}
\usepackage{multirow}
\usepackage{xr}
\usetikzlibrary{arrows.meta}

\begin{document}

\title{Predicting tensorial molecular properties with equivariant machine-learning models}

\author{Vu Ha Anh Nguyen}
\author{Alessandro Lunghi}
\email{lunghia@tcd.ie}
\affiliation{School of Physics, AMBER and CRANN Institute, Trinity College, Dublin 2, Ireland}

\begin{abstract}
{\bf Embedding molecular symmetries into machine-learning models is key for efficient learning of chemico-physical scalar properties, but little evidence on how to extend the same strategy to tensorial quantities exists. Here we formulate a scalable equivariant machine-learning model based on local atomic environment descriptors. We apply it to a series of molecules and show that accurate predictions can be achieved for a comprehensive list of dielectric and magnetic tensorial properties of different ranks. These results show that equivariant models are a promising platform to extend the scope of machine learning in materials modelling.}
\end{abstract}

\maketitle

Machine learning is revolutionizing materials science by enabling applications such as molecular properties prediction\cite{ml-matsci1,friederich2021machine}, accelerated design of new compounds\cite{ml-matsci2,ml-matsci3,ml-matsci4,ml-matsci5}, and automated AI-driven laboratories\cite{ml-matsci6}. The success of machine learning is due to its ability to understand the underlying features of a distribution of data from examples drawn from it, even in absence of physical insights on the problem\cite{early-nn1,early-nn2}. However, large improvements in the predictive power of the model are achieved by making it aware of the data's properties\cite{behler-parinello}. In the context of learning molecular properties, this is realized by including symmetries into the model's architecture. Invariance by rigid translations or swaps among identical atoms, and rotational equivariance are topical examples. 

A general physical property is described by a combination of spherical tensors $T^l_{m}(\vec{\mathbf{r}})$ of order $l$ and $2l+1$ $m$-components. When an arbitrary rotation $\hat{R}$ is applied to a molecule with coordinates $\vec{\mathbf{r}}$, the tensor rotates as
\begin{equation}
    T^l_m(\hat{R}\vec{\mathbf{r}}) = \sum_{m'} D^l_{mm'}(\hat{R})T^l_{m'}(\vec{\mathbf{r}})\:,
    \label{eq:equivariance-condition}
\end{equation}
where $D^l_{mm'}(\hat{R})$ is the Wigner $D$-matrix. Eq. \ref{eq:equivariance-condition} expresses the rotational equivariance condition that any physical property must obey. In the case of scalar quantities ($l=0$), Eq. \ref{eq:equivariance-condition} reduces to the definition of invariance. This latter scenario has been thoroughly investigated in the context of machine learning force fields\cite{ml-ff} and molecular properties prediction\cite{schnetpack,chem-shift,ml-wfn}, where atomic environments are described in terms of translationally and rotationally invariant functions\cite{behler-parinello, 4th-gen-behler, pinn, sgdml, torchani,ani1,ani2,ani3}. However,
only a few attempts at designing machine-learning models able to capture equivariance for tensorial properties so far exist.\cite{glielmo2017accurate, ceriottis-sagpr, ales-jpcc, lunghi2020insights, gauss-moment}.  

Gaussian Process Regression (GPR) has been the first machine-learning model to be applied to this problem by means of a symmetry-adapted definition of the Smooth Overlap of Atomic Position (SOAP) kernel\cite{glielmo2017accurate,ceriottis-sagpr}. In this contribution, we will instead focus on a different machine-learning architecture, with the aim of identifying the principles for equivariance learning of tensorial properties amenable to both linear and deep learning-based models. In ref. \cite{ales-jpcc}, it was proposed to extend the spectral neighbour analysis potential framework\cite{snap2,snap1}. Accordingly, a tensorial property is decomposed in a sum of atomic tensors, each one linearly dependent on the atom's bispectrum components\cite{representation}, \textit{i.e.} rotationally and translationally invariant descriptors of an atomic environment. Given the order of the tensor $l$, the equivariance was then obtained by determining $2l+1$ sets of coefficients and by reorienting any given molecular structure with respect to a reference one. Although successfully applied to the prediction of spin-phonon coupling coefficients\cite{ales-jpcc,lunghi2020insights,lunghi2020limit}, this approach is hard to generalize to multi-molecule data sets or to condensed-phase.

In this contribution, we explore the possibility to impose the equivariance condition by augmenting invariant atomic features with spherical harmonics. The latter naturally transform as in Eq. \ref{eq:equivariance-condition}, $Y^l_m(\hat{R}\vec{\mathbf{r}}) = \sum_{m'} D^l_{mm'}(\hat{R})Y^l_{m'}(\vec{\mathbf{r}})$, and therefore, provide a more natural basis for decomposing the tensorial property. This ansatz is also at the basis of equivariant models recently appeared in literature. Euclidean neural networks\cite{tensor-field, 3d-steerable, clebsch-gordan-nets, cormorant, group-equiv-conv}, moment-tensor potentials\cite{moment-tensor-potentials}, and Gaussian-moment neural networks (GMNN)\cite{gauss-moment} are a few notable examples\cite{langer2021representations}. The basic idea underlying these methods is to use contractions of high-order tensors, either spherical or Cartesian, to build invariant features. Once combined with convolutional neural networks (CNN), they have shown improved performance with respect to conventional deep learning models for the prediction of scalar quantities\cite{enn}. Interestingly, equivariant models have also been applied to the learning of Hamiltonian operators and wave-functions, which could in principle be used to predict molecular properties\cite{zhang2021equivariant,unke2021se,nigam2022equivariant}. However, despite their ideal features, these models have never been applied to the direct learning of tensorial properties. The only exception is represented by GMNN, which have only recently been applied to the learning of the magnetic anisotropy tensor\cite{gauss-moment}. 

Inspired by these works, we here demonstrate the generality and accuracy of this framework. Instead of using deep-learning architectures, we decided to use linear regression to introduce our method for a two-fold reason. On the one hand, this makes it possible to clearly individuate the principles underlying equivariant models for tensors by removing the dependency of results with respect to the model's complexity. This comes at no loss of generality, as the model can be readily scaled-up with neural networks (\textit{vide infra}). On the other hand, linear machine-learning models have shown an excellent compromise between learning rate and accuracy with respect to complex architectures\cite{snap1}, which instead perform better for large data sets\cite{nn-large-data}. \\

Following the approach of ref. \cite{ales-jpcc}, we write each component of the spherical tensor $T^{l}_m$ as a sum of atomic tensors, where $T^{l}_m(a)$ is the $a$-th atom's contribution in a system with $N_a$ atoms. The total contribution reads
\begin{equation}
    T^l_m = \sum_{a}^{N_a}T^l_m(a) = \sum_{a}^{N_a} \sum_{i}^{N_i} c_i(a)B_i(a)\bar{Y}^l_m(a)\:,
    \label{eq:snat}
\end{equation}
where $i$ runs over $N_i$ bispectrum components, $B_i(a)$, and $c_i(a)$ are coefficients that need to be determined. The use of complex or real coefficients does not change the results. $\bar{Y}^l_m(a)$ are the spherical harmonics of the $a$-th atom's environment and are defined as $\bar{Y}^l_m(a)=\sum_b^{N_b}Y^l_m(\bm{\vec{r}}_{ba})$, where $b$ runs over $N_b$ neighboring atoms within $r_{cut}$ from atom $a$, $Y^l_m$ is the standard definition of a complex spherical harmonics, and $\bm{\vec{r}}_{ba}$ are the coordinates of the atom $b$ rescaled by the coordinates of the atom $a$. The bispectrum components for each atom $a$ are also computed up to the same cutoff radius $r_{cut}$. The coefficients $c_i(a)$ are determined by minimizing the root mean squared error (RMSE) with respect to a training set of reference values $T^{l}_{m}(\vec{\mathbf{r}})$. Here we use ridge regression, which includes a $L^{2}$-regularization. The definitions of $B_i(a)$ and $\bar{Y}^l_m(a)$ enforce the model's translational invariance, while the symmetry with respect to the swap of identical atoms is imposed by using the same set of coefficients $c_{i}(a)$ for atoms of the same chemical element. The sum over atomic contributions in Eq. \ref{eq:snat} makes the model independent on the atoms' order. Finally, the model also satisfies the equivariance condition of Eq. \ref{eq:equivariance-condition}, as the right-hand side automatically transforms as a spherical tensor. This is made possible by the fact that the coefficients multiplying $\bar{Y}^{l}_{m}$ do not depend on the index $m$.

In order to illustrate the power of our model, we apply it to series of dielectric and magnetic molecular properties of key importance for various spectroscopies, namely the dipole moment, the polarizability tensor, magnetic anisotropy and the coefficients of effective crystal field Hamiltonian. The dipole moment and the polarizability tensor are needed for the modelling of infra-red and Raman spectra\cite{ir-raman, ir-raman2}, while the molecular magnetic anisotropy and the effective crystal field Hamiltonian underpin electron paramagnetic resonance (EPR) spectroscopy and spin relaxation\cite{lunghi2020limit,briganti2021,magnetic-appl2}.\\

\begin{figure}[!h]
    \centering
    \includegraphics[width=\linewidth]{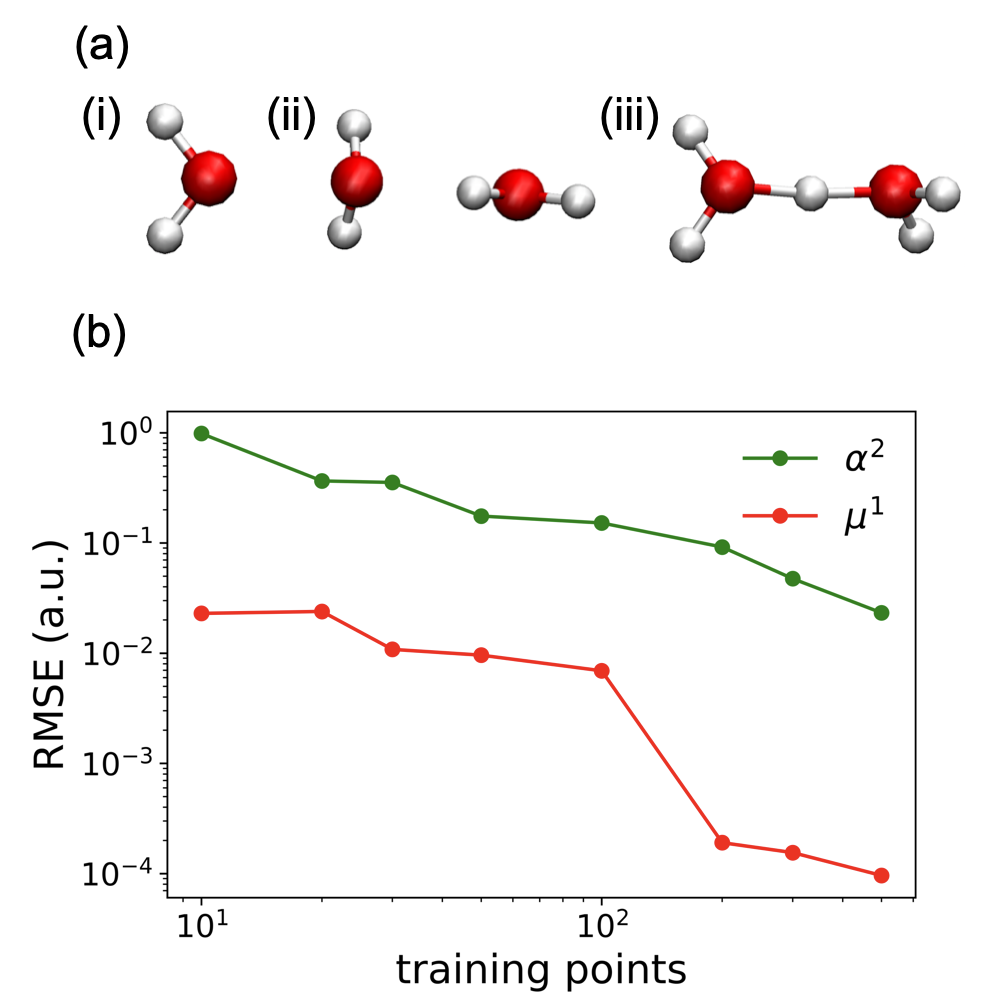}
    \caption{\textbf{Learning of dielectric properties.} (a) The molecular structure of (i) the water monomer, (ii) the water dimer, and (iii) the Zundel cation. The Oxygen atom is in red and Hydrogen atoms are in white. (b) The learning curve of our model over the water monomer data set for the polarizability spherical tensor $\alpha^2$ and the dipole spherical tensor $\mu^1$. The test set always contains 500 configurations and all quantities are expressed in atomic units.}
    \label{fig:water}
\end{figure}

\begin{figure*}[!htbp]
    \centering
    \includegraphics[width=\linewidth]{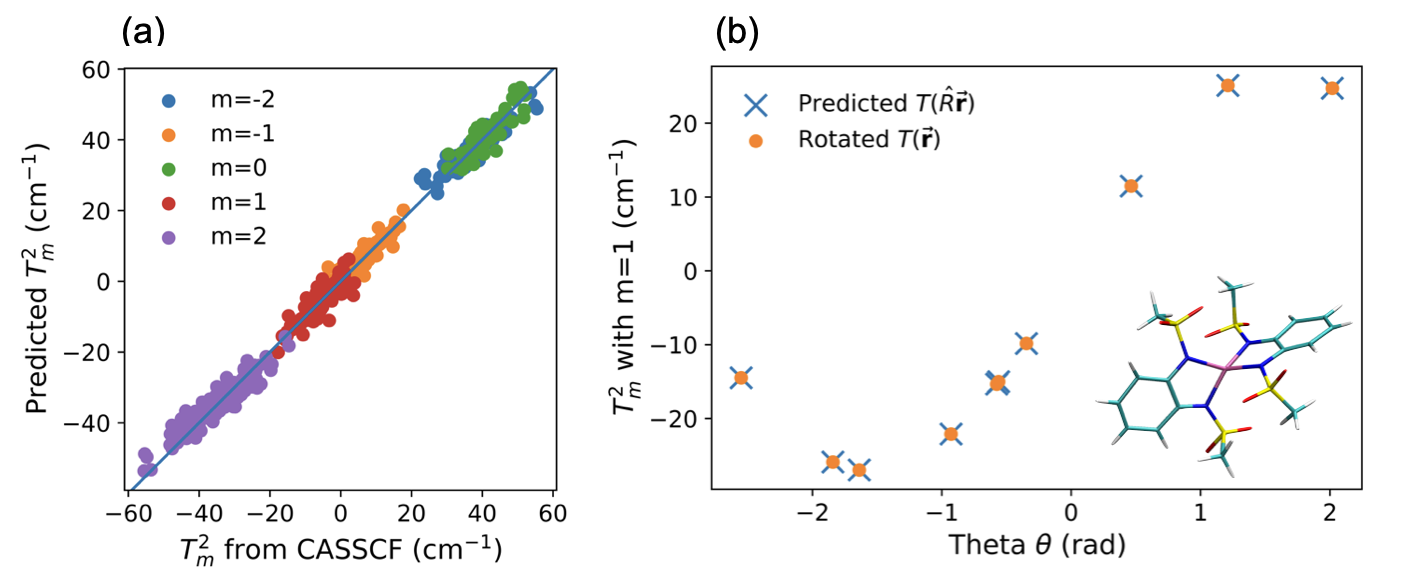}
    \caption{\textbf{Learning of magnetic anisotropy.} (a) Plot of the predicted spherical tensor's components, $T^{2}_m$, using our model against spherical tensors calculated from  the $\textbf{D}$ tensor obtained by CASSCF calculation for different configurations of CoL$_2$. (b) Plot of the predicted spherical tensor component $T^{2}_m$ with $m=1$ when the molecule is rotated by an angle $\theta$ and comparison to the rotated tensor component $T^{2}_1$ calculated by applying the Wigner $D$-matrix to the predicted tensor of the unrotated structure $\vec{\mathbf{r}}$. The inset shows the molecular structure of CoL$_2$ with the Co atom in pink, the N atoms in blue, the S atoms in yellow, the O atoms in red, the C atoms in cyan, and the H atoms in white. The plots for all other spherical tensor components are provided as Supplementary Information.}
    \label{fig:copdms}
\end{figure*}

Let us begin by discussing the results over dielectric properties for three data sets presented in ref. \cite{ceriottis-sagpr}: a water monomer, a water dimer, and the Zundel cation. The structure of these compounds is reported in Fig. \ref{fig:water}. The target properties for our model are the dipole moment vector $\bm{\mu}$, which is equivalent to a spherical tensor with $l=1$ ($\mu^1$), and the polarizability tensor $\bm{\alpha}$, which is equivalent to the sum of two spherical tensors with $l=0$ and $l=2$, respectively. Here we focus on learning the non-scalar component with $l=2$ ($\alpha^2$). The three data sets each contains 1000 configurations with arbitrary distortion and orientation in space\cite{ceriottis-sagpr}. Both the dipole and the polarizability tensor were computed at the CCSD/d-aug-cc-pvtz level. Fig. \ref{fig:water}b shows the learning curve our model for the water monomer data set, namely the plot of the test set's RMSE as function of the number of training points. The learning curve for water dimer and Zundel cation data sets show similar trends and they are reported in Fig. S1. It is clear that with more training points, the test error goes down and the accuracy of the model's prediction improves. Although a thorough comparison of different methods is beyond the scope of this work, we found that symmetry-adapted Gaussian process regression (SA-GPR)\cite{ceriottis-sagpr} performs similarly or slightly better than our linear model (see Table S1). \\

Next, we apply our model to the magnetic properties of coordination compounds. The target property for the following tests is the magnetic anisotropy tensor \textbf{D}, which can be accurately computed using Complete Active Space (CASSCF) and valence state perturbation theory (NEVPT2) methods\cite{neese2019chemistry}. \textbf{D} is a symmetric trace-less tensor and can thus be converted into a spherical tensors of order $l=2$.\\

Firstly, we study a data set that contains only local distortions of a single molecular structure, namely the one of the top-performance single molecule magnet CoL$_2$ (with H$_{2}$L=1,2-bis-
(methanesulfonamido)benzene)\cite{copdms}. The data set, presented in ref. \cite{ales-jpcc}, contains at total of 1500 configurations obtained by applying a random distortion of maximum displacements of $\pm 0.05$ \AA$ $ to each atom's Cartesian components. 1200 configurations were used to train the model, 150 were used as validation to tune the hyper-parameters, and 150 were used as test set. Fig. \ref{fig:copdms}a illustrates the accuracy of our model in predicting the magnetic anisotropy, with an error of 1.6 cm$^{-1}$ for all $T^{2}_m$. Most importantly, it should be noted that the orientation of the configurations in the training and test sets does not affect the accuracy of the model. The error of the model remains identical when the configurations in the data set are all oriented in the same direction or when the configurations are randomly rotated. To further demonstrate that our method is in fact equivariant, a configuration $\vec{\mathbf{r}}$ of the test set is randomly rotated by an angle $\theta$. The predicted spherical tensor for this configuration $T^{2}_m(\vec{\mathbf{r}})$ is rotated using the corresponding Wigner $D$-matrix. Then, our model is used to predict the tensor values for the rotated configuration $T^{2}_m(\hat{R}\vec{\mathbf{r}})$. Based on Eq. \ref{eq:equivariance-condition}, these two values should be identical for an equivariant model, which is indeed the case (see Fig. \ref{fig:copdms}b and Fig. S8). 

\begin{figure}[t]
    \centering
    \includegraphics[width=\linewidth]{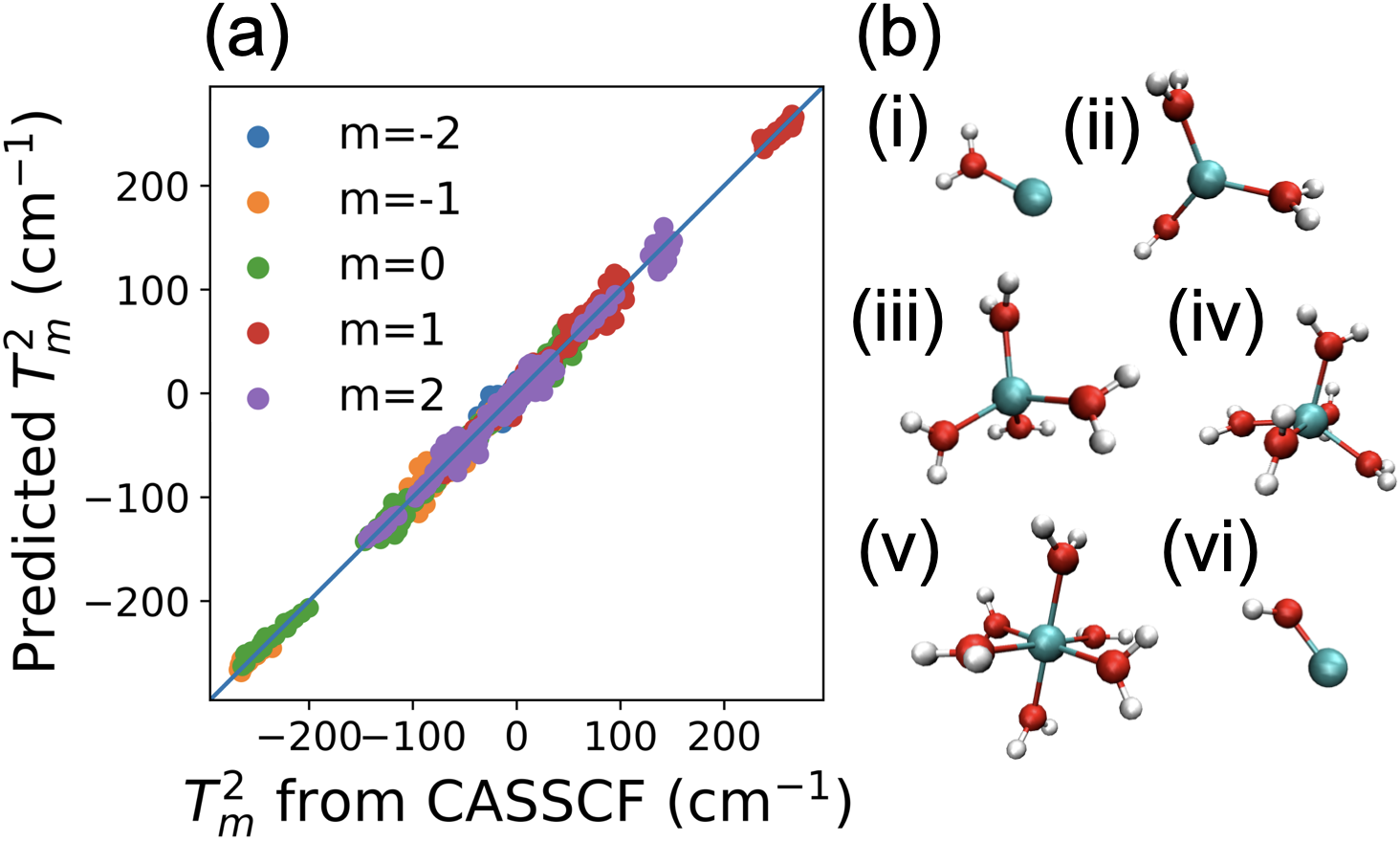}
    \caption{(a) Plot of predicted spherical tensors $T^{2}_m$ using our model against spherical tensors calculated from CASSCF $\textbf{D}$ tensor values of the Co(H$_2$O)$_x$ set. (b) The molecular structure of (i) [Co(H$_2$O)$_1$]$^{2+}$, (ii) [Co(H$_2$O)$_3$]$^{2+}$, (iii) [Co(H$_2$O)$_4$]$^{2+}$, (iv) [Co(H$_2$O)$_5$]$^{2+}$, (v) [Co(H$_2$O)$_6$]$^{2+}$, and (vi) [Co(OH)$_1$]$^{+}$ where the Co atom is coloured in cyan, the O atoms are coloured in red, and the H atoms are coloured in white.}
    \label{fig:coh2ox}
\end{figure}

Next we apply our model to a data set that contains multiple molecules. The data set includes the six molecules [Co(H$_2$O)$_1$]$^{2+}$, [Co(H$_2$O)$_3$]$^{2+}$, [Co(H$_2$O)$_4$]$^{2+}$, [Co(H$_2$O)$_5$]$^{2+}$, [Co(H$_2$O)$_6$]$^{2+}$, and [Co(OH)$_1$]$^{+}$. Each molecule was randomly distorted 100 times by applying a maximum displacements of $\pm 0.05$ \AA$ $ to each atom's Cartesian components. 80 configurations per molecules were used for training and 20 configurations per molecule as test set. The tensor \textbf{D} was then computed with CASSCF for each molecular frame. The accuracy of our predictions is illustrated in Fig. \ref{fig:coh2ox}a, with an error of 5.9 cm$^{-1}$ for all $m$ components. Similarly to the previous test with CoL$_2$, the accuracy of the model does not depend on the specific orientation of the molecules.
To further test the reliability of the model's predictions, we use it to predict how the tensor $\mathbf{D}$ changes as function of small distortions of the Co$^{2+}$ ion in [Co(H$_2$O)$_6$]$^{2+}$. Although none of these configurations explicitly appear in the training set, Fig. S4 shows that the correct profile of $T^{2}_m$ as a function of the atomic displacement is well reproduced. The positive outcome of this test enables applications in spin relaxation, where the derivatives of $\mathbf{D}(\vec{\mathbf{r}})$ are needed\cite{lunghi2020limit,lunghi2020multiple}. As a second test, we re-train the model by adding an additional spectator water molecule at great distance from the Co complexes. We compute the contribution to the total axial magnetic anisotropy $|D|$ for a water molecule bonded to one of the Co(H$_2$O)$_x$ molecules and for the spectator one. Fig. S3 shows that the model correctly captures the physics of the problem by assigning a vanishing $|D|$ value, within the model's RMSE, to the non-bonded water molecule, and a finite $|D|$ value to the water molecules coordinating the Co ion.\\
\begin{figure}[t]
    \centering
    \includegraphics[width=\linewidth]{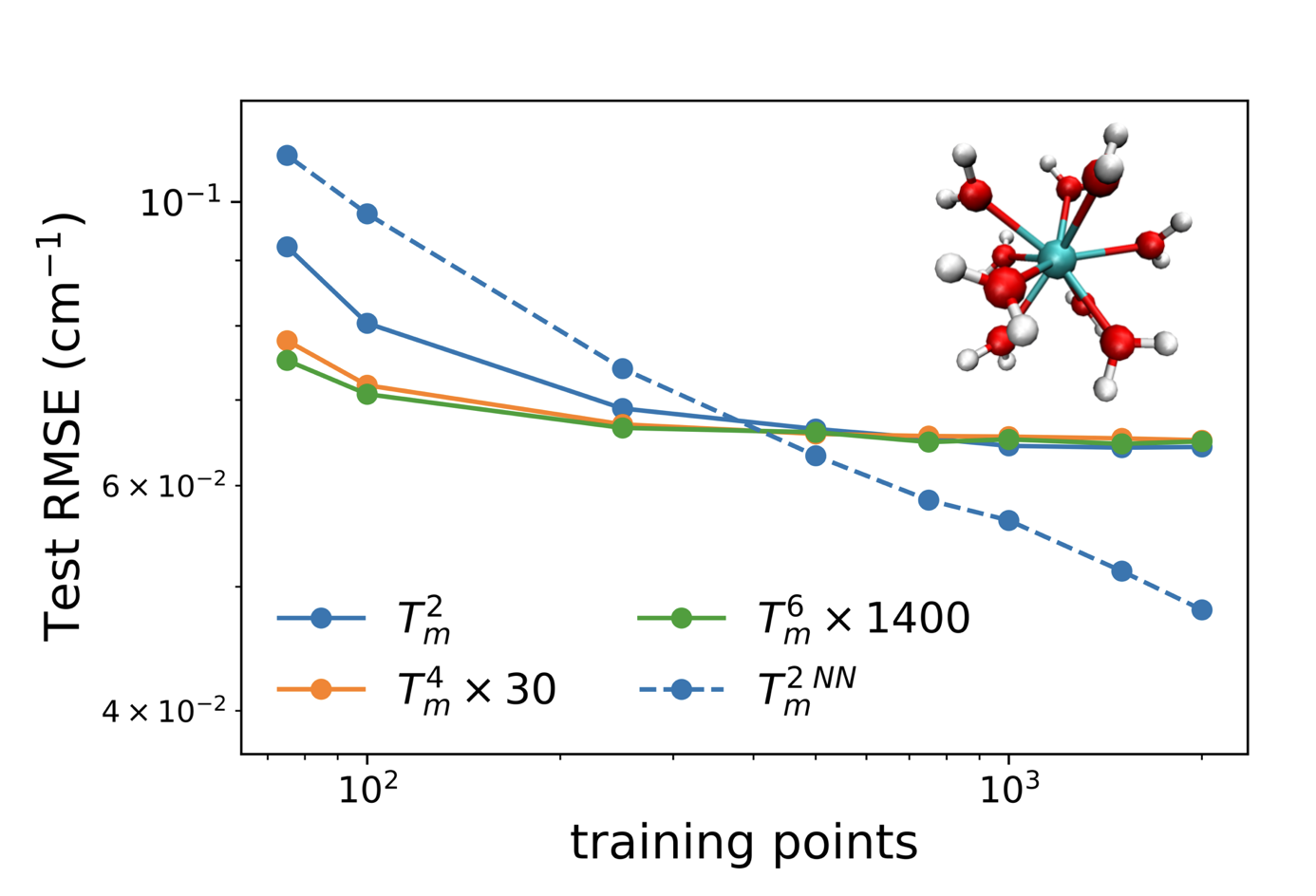}
    \caption{\textbf{Learning of crystal field parameters.} The learning curve of our model over the [Dy(H$_2$O)$_9$]$^{3+}$ data set for the spherical tensors of crystal field parameters of order $l=2,4,6$. The test set always contains 430 configurations. Continuous lines and filled circles refer to the results of the linear model, dashed lines and filled circles to the results obtained with a neural network architecture. The inset shows the molecular structure of [Dy(H$_2$O)$_9$]$^{3+}$ where the Dy atom is in cyan, the O atoms are in red, and H atoms are in white.}
    \label{fig:learning-curve-dyh2o}
\end{figure}

Next we present the results for the magnetic properties of the [Dy(H$_{2}$O)$_{9}$]$^{3+}$ coordination complex, whose structure is depicted in the inset of Fig. \ref{fig:learning-curve-dyh2o}. The Dy$^{3+}$ ion posses 5 unpaired electrons in the $f$-shell and an un-quenched electronic angular momentum, resulting in a ground state with a total angular moment $\vec{\mathbf{J}}=15/2$. The 16 electronic levels associated with this multiplet can be described with the effective crystal field Hamiltonian
\begin{equation}
    \hat{H}_{CF} = \sum_{l=2,4,6}\sum_{m=-l}^{l} B^{l}_{m}\hat{O}_{m}^{l}(\vec{\mathbf{J}}) \:,
    \label{eq:cfham}
\end{equation}
where $B^{l}_{m}$ are coefficients adjusted to reproduce CASSCF simulations\cite{ungur-cf}. The operators $\hat{O}_{m}^{l}(\vec{\mathbf{J}})$ are tesseral tensor operators, which can be transformed into spherical tensor operators\cite{ryabov-1999}. 
We prepare a training set of over 3000 entries by distorting the DFT-optimized structure of Dy[H$_{2}$O$_{9}$]$^{3+}$. Random distortions of a maximum value of $\pm 0.05$ \AA$ $ are applied to each atom's Cartesian coordinates. The crystal field coefficients appearing in Eq. \ref{eq:cfham} are then computed with CASSCF methods as detailed in the ESI. We apply our method to the coefficients $B^{l}_{m}$, converted into the spherical components, and report the learning curve in Fig. \ref{fig:learning-curve-dyh2o}, demonstrating that tensors up to the sixth order can be well reproduced. Graphical representation of predictions vs reference data and proof of model equivariance for this data set are reported in Figs. S5 and S9-S11.

Finally, we use the Dy[H$_{2}$O$_{9}$]$^{3+}$ data sets to also show that the model of Eq. \ref{eq:snat} can be easily generalized to deep learning architectures by feeding the bispectrum components, $\{B(a)\}$, to a neural network with  weights/biases $\{c(a)\}$, and real scalar output $f$ (see Fig. S6)
\begin{equation}
    T^l_m = \sum_{a}^{N_a}T^l_m(a) = \sum_{a}^{N_a}  f_{a}\left[ \{c(a)\},\{B(a)\}\right]\bar{Y}^l_m(a)\:.
    \label{eq:snat2}
\end{equation}\\
Results for a model with one independent neural net for each chemical species, each with three hidden layers and 32:32:16 nodes, are reported in Fig. \ref{fig:learning-curve-dyh2o} and Fig. S7 and highlight the fact that the linear model provides better results for a small training set. However, as expected, as larger training sets are employed, the deep-learning model is able to keep leaning from data, while the performance of the linear model reaches a plateau due to the limited flexibility of its architecture. Similar results are obtained for tensors of higher order and for neural networks with different numbers of layers or nodes (see ESI). The excellent results obtained with Eq. \ref{eq:snat2} demonstrate the generality of our approach, where models designed to predict scalar properties can be extended to the case of tensors by simply learning the coefficients of a combination of spherical harmonics functions. \\

In conclusion, we have presented a general procedure to build equivariant machine-learning models able to predict tensorial properties. Even in its simple linear form, the model accurately predicts properties of different type and rank, and it is able to handle realistic compounds as well as multi-molecule data sets. Moreover, we have shown that higher accuracy can be achieved be using deep-learning architectures and large data sets. This model can find immediate application in magnetism\cite{magnetic-appl2,ml-mag-bulk}, magnetic resonance\cite{paruzzo2018chemical,lunghi2020insights}, and vibrational spectroscopy\cite{ir-raman2}, or be further optimized by combining it with the wealth of methods already developed for the prediction of scalar properties\cite{behler-parinello,ml-matsci1,friederich2021machine,schnetpack}.\\

\newpage

\vspace{0.2cm}
\noindent
\textbf{Acknowledgements}\\
This project has received funding from the European Research Council (ERC) under the European Union’s Horizon 2020 research and innovation programme (grant agreement No. [948493]). Computational resources were provided by the Trinity College Research IT and the Irish Centre for High-End Computing (ICHEC).

%


\begin{thebibliography}{53}%
\makeatletter
\providecommand \@ifxundefined [1]{%
 \@ifx{#1\undefined}
}%
\providecommand \@ifnum [1]{%
 \ifnum #1\expandafter \@firstoftwo
 \else \expandafter \@secondoftwo
 \fi
}%
\providecommand \@ifx [1]{%
 \ifx #1\expandafter \@firstoftwo
 \else \expandafter \@secondoftwo
 \fi
}%
\providecommand \natexlab [1]{#1}%
\providecommand \enquote  [1]{``#1''}%
\providecommand \bibnamefont  [1]{#1}%
\providecommand \bibfnamefont [1]{#1}%
\providecommand \citenamefont [1]{#1}%
\providecommand \href@noop [0]{\@secondoftwo}%
\providecommand \href [0]{\begingroup \@sanitize@url \@href}%
\providecommand \@href[1]{\@@startlink{#1}\@@href}%
\providecommand \@@href[1]{\endgroup#1\@@endlink}%
\providecommand \@sanitize@url [0]{\catcode `\\12\catcode `\$12\catcode
  `\&12\catcode `\#12\catcode `\^12\catcode `\_12\catcode `\%12\relax}%
\providecommand \@@startlink[1]{}%
\providecommand \@@endlink[0]{}%
\providecommand \url  [0]{\begingroup\@sanitize@url \@url }%
\providecommand \@url [1]{\endgroup\@href {#1}{\urlprefix }}%
\providecommand \urlprefix  [0]{URL }%
\providecommand \Eprint [0]{\href }%
\providecommand \doibase [0]{http://dx.doi.org/}%
\providecommand \selectlanguage [0]{\@gobble}%
\providecommand \bibinfo  [0]{\@secondoftwo}%
\providecommand \bibfield  [0]{\@secondoftwo}%
\providecommand \translation [1]{[#1]}%
\providecommand \BibitemOpen [0]{}%
\providecommand \bibitemStop [0]{}%
\providecommand \bibitemNoStop [0]{.\EOS\space}%
\providecommand \EOS [0]{\spacefactor3000\relax}%
\providecommand \BibitemShut  [1]{\csname bibitem#1\endcsname}%
\let\auto@bib@innerbib\@empty
\bibitem [{\citenamefont {Butler}\ \emph {et~al.}(2018)\citenamefont {Butler},
  \citenamefont {Davies}, \citenamefont {Cartwright}, \citenamefont {Isayev},\
  and\ \citenamefont {Walsh}}]{ml-matsci1}%
  \BibitemOpen
  \bibfield  {author} {\bibinfo {author} {\bibfnamefont {K.~T.}\ \bibnamefont
  {Butler}}, \bibinfo {author} {\bibfnamefont {D.~W.}\ \bibnamefont {Davies}},
  \bibinfo {author} {\bibfnamefont {H.}~\bibnamefont {Cartwright}}, \bibinfo
  {author} {\bibfnamefont {O.}~\bibnamefont {Isayev}}, \ and\ \bibinfo {author}
  {\bibfnamefont {A.}~\bibnamefont {Walsh}},\ }\href@noop {} {\bibfield
  {journal} {\bibinfo  {journal} {Nature}\ }\textbf {\bibinfo {volume} {559}},\
  \bibinfo {pages} {547–555} (\bibinfo {year} {2018})}\BibitemShut {NoStop}%
\bibitem [{\citenamefont {Friederich}\ \emph {et~al.}(2021)\citenamefont
  {Friederich}, \citenamefont {H{\"a}se}, \citenamefont {Proppe},\ and\
  \citenamefont {Aspuru-Guzik}}]{friederich2021machine}%
  \BibitemOpen
  \bibfield  {author} {\bibinfo {author} {\bibfnamefont {P.}~\bibnamefont
  {Friederich}}, \bibinfo {author} {\bibfnamefont {F.}~\bibnamefont
  {H{\"a}se}}, \bibinfo {author} {\bibfnamefont {J.}~\bibnamefont {Proppe}}, \
  and\ \bibinfo {author} {\bibfnamefont {A.}~\bibnamefont {Aspuru-Guzik}},\
  }\href@noop {} {\bibfield  {journal} {\bibinfo  {journal} {Nat. Mater.}\
  }\textbf {\bibinfo {volume} {20}},\ \bibinfo {pages} {750} (\bibinfo {year}
  {2021})}\BibitemShut {NoStop}%
\bibitem [{\citenamefont {Raccuglia}\ \emph {et~al.}(2016)\citenamefont
  {Raccuglia}, \citenamefont {Elbert}, \citenamefont {Adler}, \citenamefont
  {Falk}, \citenamefont {Wenny}, \citenamefont {Mollo}, \citenamefont {Zeller},
  \citenamefont {Friedler}, \citenamefont {Schrier},\ and\ \citenamefont
  {Norquist}}]{ml-matsci2}%
  \BibitemOpen
  \bibfield  {author} {\bibinfo {author} {\bibfnamefont {P.}~\bibnamefont
  {Raccuglia}}, \bibinfo {author} {\bibfnamefont {K.~C.}\ \bibnamefont
  {Elbert}}, \bibinfo {author} {\bibfnamefont {P.~D.~F.}\ \bibnamefont
  {Adler}}, \bibinfo {author} {\bibfnamefont {C.}~\bibnamefont {Falk}},
  \bibinfo {author} {\bibfnamefont {M.~B.}\ \bibnamefont {Wenny}}, \bibinfo
  {author} {\bibfnamefont {A.}~\bibnamefont {Mollo}}, \bibinfo {author}
  {\bibfnamefont {M.}~\bibnamefont {Zeller}}, \bibinfo {author} {\bibfnamefont
  {S.~A.}\ \bibnamefont {Friedler}}, \bibinfo {author} {\bibfnamefont
  {J.}~\bibnamefont {Schrier}}, \ and\ \bibinfo {author} {\bibfnamefont
  {A.~J.}\ \bibnamefont {Norquist}},\ }\href@noop {} {\bibfield  {journal}
  {\bibinfo  {journal} {Nature}\ }\textbf {\bibinfo {volume} {533}},\ \bibinfo
  {pages} {73–76} (\bibinfo {year} {2016})}\BibitemShut {NoStop}%
\bibitem [{\citenamefont {Saal}\ \emph {et~al.}(2020)\citenamefont {Saal},
  \citenamefont {Oliynyk},\ and\ \citenamefont {Meredig}}]{ml-matsci3}%
  \BibitemOpen
  \bibfield  {author} {\bibinfo {author} {\bibfnamefont {J.~E.}\ \bibnamefont
  {Saal}}, \bibinfo {author} {\bibfnamefont {A.~O.}\ \bibnamefont {Oliynyk}}, \
  and\ \bibinfo {author} {\bibfnamefont {B.}~\bibnamefont {Meredig}},\
  }\href@noop {} {\bibfield  {journal} {\bibinfo  {journal} {Annu. Rev. Mater.
  Res.}\ }\textbf {\bibinfo {volume} {50}},\ \bibinfo {pages} {49} (\bibinfo
  {year} {2020})}\BibitemShut {NoStop}%
\bibitem [{\citenamefont {Suh}\ \emph {et~al.}(2020)\citenamefont {Suh},
  \citenamefont {Fare}, \citenamefont {Warren},\ and\ \citenamefont
  {Pyzer-Knapp}}]{ml-matsci4}%
  \BibitemOpen
  \bibfield  {author} {\bibinfo {author} {\bibfnamefont {C.}~\bibnamefont
  {Suh}}, \bibinfo {author} {\bibfnamefont {C.}~\bibnamefont {Fare}}, \bibinfo
  {author} {\bibfnamefont {J.~A.}\ \bibnamefont {Warren}}, \ and\ \bibinfo
  {author} {\bibfnamefont {E.~O.}\ \bibnamefont {Pyzer-Knapp}},\ }\href@noop {}
  {\bibfield  {journal} {\bibinfo  {journal} {Annu. Rev. Mater. Res.}\ }\textbf
  {\bibinfo {volume} {50}},\ \bibinfo {pages} {1} (\bibinfo {year}
  {2020})}\BibitemShut {NoStop}%
\bibitem [{\citenamefont {Zhang}\ \emph {et~al.}(2018)\citenamefont {Zhang},
  \citenamefont {Hippalgaonkar}, \citenamefont {Buonassisi}, \citenamefont
  {Lovvik}, \citenamefont {Sagvolden}, \citenamefont {Ding},\ and\
  \citenamefont {et~al.}}]{ml-matsci5}%
  \BibitemOpen
  \bibfield  {author} {\bibinfo {author} {\bibfnamefont {H.}~\bibnamefont
  {Zhang}}, \bibinfo {author} {\bibfnamefont {K.}~\bibnamefont
  {Hippalgaonkar}}, \bibinfo {author} {\bibfnamefont {T.}~\bibnamefont
  {Buonassisi}}, \bibinfo {author} {\bibfnamefont {O.~M.}\ \bibnamefont
  {Lovvik}}, \bibinfo {author} {\bibfnamefont {E.}~\bibnamefont {Sagvolden}},
  \bibinfo {author} {\bibfnamefont {D.}~\bibnamefont {Ding}}, \ and\ \bibinfo
  {author} {\bibnamefont {et~al.}},\ }\href@noop {} {\bibfield  {journal}
  {\bibinfo  {journal} {ES Energy \& Environment}\ } (\bibinfo {year}
  {2018})}\BibitemShut {NoStop}%
\bibitem [{\citenamefont {de~Almeida}\ \emph {et~al.}(2019)\citenamefont
  {de~Almeida}, \citenamefont {Moreira},\ and\ \citenamefont
  {Rodrigues}}]{ml-matsci6}%
  \BibitemOpen
  \bibfield  {author} {\bibinfo {author} {\bibfnamefont {A.~F.}\ \bibnamefont
  {de~Almeida}}, \bibinfo {author} {\bibfnamefont {R.}~\bibnamefont {Moreira}},
  \ and\ \bibinfo {author} {\bibfnamefont {T.}~\bibnamefont {Rodrigues}},\
  }\href@noop {} {\bibfield  {journal} {\bibinfo  {journal} {Nat. Rev. Chem.}\
  }\textbf {\bibinfo {volume} {3}},\ \bibinfo {pages} {589–604} (\bibinfo
  {year} {2019})}\BibitemShut {NoStop}%
\bibitem [{\citenamefont {de~Almeida}\ \emph {et~al.}(1995)\citenamefont
  {de~Almeida}, \citenamefont {Moreira},\ and\ \citenamefont
  {Rodrigues}}]{early-nn1}%
  \BibitemOpen
  \bibfield  {author} {\bibinfo {author} {\bibfnamefont {A.~F.}\ \bibnamefont
  {de~Almeida}}, \bibinfo {author} {\bibfnamefont {R.}~\bibnamefont {Moreira}},
  \ and\ \bibinfo {author} {\bibfnamefont {T.}~\bibnamefont {Rodrigues}},\
  }\href@noop {} {\bibfield  {journal} {\bibinfo  {journal} {J. Chem. Phys.}\
  }\textbf {\bibinfo {volume} {103}},\ \bibinfo {pages} {4129} (\bibinfo {year}
  {1995})}\BibitemShut {NoStop}%
\bibitem [{\citenamefont {Lorenz}\ \emph {et~al.}(2004)\citenamefont {Lorenz},
  \citenamefont {Gro\ss},\ and\ \citenamefont {Scheffler}}]{early-nn2}%
  \BibitemOpen
  \bibfield  {author} {\bibinfo {author} {\bibfnamefont {S.}~\bibnamefont
  {Lorenz}}, \bibinfo {author} {\bibfnamefont {A.}~\bibnamefont {Gro\ss}}, \
  and\ \bibinfo {author} {\bibfnamefont {M.}~\bibnamefont {Scheffler}},\
  }\href@noop {} {\bibfield  {journal} {\bibinfo  {journal} {Chem. Phys.
  Lett.}\ }\textbf {\bibinfo {volume} {395}},\ \bibinfo {pages} {210} (\bibinfo
  {year} {2004})}\BibitemShut {NoStop}%
\bibitem [{\citenamefont {Behler}\ and\ \citenamefont
  {Parrinello}(2007)}]{behler-parinello}%
  \BibitemOpen
  \bibfield  {author} {\bibinfo {author} {\bibfnamefont {J.}~\bibnamefont
  {Behler}}\ and\ \bibinfo {author} {\bibfnamefont {M.}~\bibnamefont
  {Parrinello}},\ }\href@noop {} {\bibfield  {journal} {\bibinfo  {journal}
  {Phys. Rev. Lett.}\ }\textbf {\bibinfo {volume} {98}} (\bibinfo {year}
  {2007})}\BibitemShut {NoStop}%
\bibitem [{\citenamefont {Poltavsky}\ and\ \citenamefont
  {Tkatchenko}(2021)}]{ml-ff}%
  \BibitemOpen
  \bibfield  {author} {\bibinfo {author} {\bibfnamefont {I.}~\bibnamefont
  {Poltavsky}}\ and\ \bibinfo {author} {\bibfnamefont {A.}~\bibnamefont
  {Tkatchenko}},\ }\href@noop {} {\bibfield  {journal} {\bibinfo  {journal} {J.
  Phys. Chem. Lett.}\ }\textbf {\bibinfo {volume} {12}},\ \bibinfo {pages}
  {6551–6564} (\bibinfo {year} {2021})}\BibitemShut {NoStop}%
\bibitem [{\citenamefont {Sch\"{u}tt}\ \emph
  {et~al.}(2019{\natexlab{a}})\citenamefont {Sch\"{u}tt}, \citenamefont
  {Kessel}, \citenamefont {Gastegger}, \citenamefont {Nicoli}, \citenamefont
  {Tkatchenko},\ and\ \citenamefont {M\"{u}ller}}]{schnetpack}%
  \BibitemOpen
  \bibfield  {author} {\bibinfo {author} {\bibfnamefont {K.~T.}\ \bibnamefont
  {Sch\"{u}tt}}, \bibinfo {author} {\bibfnamefont {P.}~\bibnamefont {Kessel}},
  \bibinfo {author} {\bibfnamefont {M.}~\bibnamefont {Gastegger}}, \bibinfo
  {author} {\bibfnamefont {K.~A.}\ \bibnamefont {Nicoli}}, \bibinfo {author}
  {\bibfnamefont {A.}~\bibnamefont {Tkatchenko}}, \ and\ \bibinfo {author}
  {\bibfnamefont {K.-R.}\ \bibnamefont {M\"{u}ller}},\ }\href@noop {}
  {\bibfield  {journal} {\bibinfo  {journal} {J. Chem. Theory Comput.}\
  }\textbf {\bibinfo {volume} {15}},\ \bibinfo {pages} {448} (\bibinfo {year}
  {2019}{\natexlab{a}})}\BibitemShut {NoStop}%
\bibitem [{\citenamefont {Paruzzo}\ \emph
  {et~al.}(2018{\natexlab{a}})\citenamefont {Paruzzo}, \citenamefont
  {Hofstetter}, \citenamefont {Musil}, \citenamefont {De}, \citenamefont
  {Ceriotti},\ and\ \citenamefont {Emsley}}]{chem-shift}%
  \BibitemOpen
  \bibfield  {author} {\bibinfo {author} {\bibfnamefont {F.~M.}\ \bibnamefont
  {Paruzzo}}, \bibinfo {author} {\bibfnamefont {A.}~\bibnamefont {Hofstetter}},
  \bibinfo {author} {\bibfnamefont {F.}~\bibnamefont {Musil}}, \bibinfo
  {author} {\bibfnamefont {S.}~\bibnamefont {De}}, \bibinfo {author}
  {\bibfnamefont {M.}~\bibnamefont {Ceriotti}}, \ and\ \bibinfo {author}
  {\bibfnamefont {L.}~\bibnamefont {Emsley}},\ }\href@noop {} {\bibfield
  {journal} {\bibinfo  {journal} {Nat. Commun.}\ }\textbf {\bibinfo {volume}
  {9}},\ \bibinfo {pages} {4501} (\bibinfo {year}
  {2018}{\natexlab{a}})}\BibitemShut {NoStop}%
\bibitem [{\citenamefont {Sch\"{u}tt}\ \emph
  {et~al.}(2019{\natexlab{b}})\citenamefont {Sch\"{u}tt}, \citenamefont
  {Gastegger}, \citenamefont {Tkatchenko}, \citenamefont {M\"{u}ller},\ and\
  \citenamefont {Maurer}}]{ml-wfn}%
  \BibitemOpen
  \bibfield  {author} {\bibinfo {author} {\bibfnamefont {K.~T.}\ \bibnamefont
  {Sch\"{u}tt}}, \bibinfo {author} {\bibfnamefont {M.}~\bibnamefont
  {Gastegger}}, \bibinfo {author} {\bibfnamefont {A.}~\bibnamefont
  {Tkatchenko}}, \bibinfo {author} {\bibfnamefont {K.-R.}\ \bibnamefont
  {M\"{u}ller}}, \ and\ \bibinfo {author} {\bibfnamefont {R.~J.}\ \bibnamefont
  {Maurer}},\ }\href@noop {} {\bibfield  {journal} {\bibinfo  {journal} {Nat.
  Commun.}\ }\textbf {\bibinfo {volume} {10}},\ \bibinfo {pages} {5024}
  (\bibinfo {year} {2019}{\natexlab{b}})}\BibitemShut {NoStop}%
\bibitem [{\citenamefont {Ko}\ \emph {et~al.}(2021)\citenamefont {Ko},
  \citenamefont {Finkler}, \citenamefont {Geodecker},\ and\ \citenamefont
  {Behler}}]{4th-gen-behler}%
  \BibitemOpen
  \bibfield  {author} {\bibinfo {author} {\bibfnamefont {T.~W.}\ \bibnamefont
  {Ko}}, \bibinfo {author} {\bibfnamefont {J.~A.}\ \bibnamefont {Finkler}},
  \bibinfo {author} {\bibfnamefont {S.}~\bibnamefont {Geodecker}}, \ and\
  \bibinfo {author} {\bibfnamefont {J.}~\bibnamefont {Behler}},\ }\href@noop {}
  {\bibfield  {journal} {\bibinfo  {journal} {Nat. Commun.}\ }\textbf {\bibinfo
  {volume} {12}} (\bibinfo {year} {2021})}\BibitemShut {NoStop}%
\bibitem [{\citenamefont {Shao}\ \emph {et~al.}(2020)\citenamefont {Shao},
  \citenamefont {Hellstr\"{o}m}, \citenamefont {Mitev}, \citenamefont
  {Knijff},\ and\ \citenamefont {Zhang}}]{pinn}%
  \BibitemOpen
  \bibfield  {author} {\bibinfo {author} {\bibfnamefont {Y.}~\bibnamefont
  {Shao}}, \bibinfo {author} {\bibfnamefont {M.}~\bibnamefont {Hellstr\"{o}m}},
  \bibinfo {author} {\bibfnamefont {P.~D.}\ \bibnamefont {Mitev}}, \bibinfo
  {author} {\bibfnamefont {L.}~\bibnamefont {Knijff}}, \ and\ \bibinfo {author}
  {\bibfnamefont {C.}~\bibnamefont {Zhang}},\ }\href@noop {} {\bibfield
  {journal} {\bibinfo  {journal} {J. Chem. Inf. Model.}\ }\textbf {\bibinfo
  {volume} {60}},\ \bibinfo {pages} {1184} (\bibinfo {year}
  {2020})}\BibitemShut {NoStop}%
\bibitem [{\citenamefont {Chmiela}\ \emph {et~al.}(2019)\citenamefont
  {Chmiela}, \citenamefont {Sauceda}, \citenamefont {Poltavsky}, \citenamefont
  {M\"{u}ller},\ and\ \citenamefont {Tkatchenko}}]{sgdml}%
  \BibitemOpen
  \bibfield  {author} {\bibinfo {author} {\bibfnamefont {S.}~\bibnamefont
  {Chmiela}}, \bibinfo {author} {\bibfnamefont {H.~E.}\ \bibnamefont
  {Sauceda}}, \bibinfo {author} {\bibfnamefont {I.}~\bibnamefont {Poltavsky}},
  \bibinfo {author} {\bibfnamefont {K.-R.}\ \bibnamefont {M\"{u}ller}}, \ and\
  \bibinfo {author} {\bibfnamefont {A.}~\bibnamefont {Tkatchenko}},\
  }\href@noop {} {\bibfield  {journal} {\bibinfo  {journal} {Comput. Phys.
  Commun.}\ }\textbf {\bibinfo {volume} {240}},\ \bibinfo {pages} {38}
  (\bibinfo {year} {2019})}\BibitemShut {NoStop}%
\bibitem [{\citenamefont {Gao}\ \emph {et~al.}(2020)\citenamefont {Gao},
  \citenamefont {Ramezanghorbani}, \citenamefont {Isayev}, \citenamefont
  {Smith},\ and\ \citenamefont {Roitberg}}]{torchani}%
  \BibitemOpen
  \bibfield  {author} {\bibinfo {author} {\bibfnamefont {X.}~\bibnamefont
  {Gao}}, \bibinfo {author} {\bibfnamefont {F.}~\bibnamefont
  {Ramezanghorbani}}, \bibinfo {author} {\bibfnamefont {O.}~\bibnamefont
  {Isayev}}, \bibinfo {author} {\bibfnamefont {J.~S.}\ \bibnamefont {Smith}}, \
  and\ \bibinfo {author} {\bibfnamefont {A.~E.}\ \bibnamefont {Roitberg}},\
  }\href@noop {} {\bibfield  {journal} {\bibinfo  {journal} {J. Chem. Inf.
  Model}\ }\textbf {\bibinfo {volume} {60}},\ \bibinfo {pages} {3408} (\bibinfo
  {year} {2020})}\BibitemShut {NoStop}%
\bibitem [{\citenamefont {Devereux}\ \emph {et~al.}(2020)\citenamefont
  {Devereux}, \citenamefont {Smith}, \citenamefont {Huddleston}, \citenamefont
  {Barros}, \citenamefont {Zubatyuk}, \citenamefont {Isayev},\ and\
  \citenamefont {Roitberg}}]{ani1}%
  \BibitemOpen
  \bibfield  {author} {\bibinfo {author} {\bibfnamefont {C.}~\bibnamefont
  {Devereux}}, \bibinfo {author} {\bibfnamefont {J.~S.}\ \bibnamefont {Smith}},
  \bibinfo {author} {\bibfnamefont {K.~K.}\ \bibnamefont {Huddleston}},
  \bibinfo {author} {\bibfnamefont {K.}~\bibnamefont {Barros}}, \bibinfo
  {author} {\bibfnamefont {R.}~\bibnamefont {Zubatyuk}}, \bibinfo {author}
  {\bibfnamefont {O.}~\bibnamefont {Isayev}}, \ and\ \bibinfo {author}
  {\bibfnamefont {A.~E.}\ \bibnamefont {Roitberg}},\ }\href@noop {} {\bibfield
  {journal} {\bibinfo  {journal} {J. Chem. Theory Comput.}\ }\textbf {\bibinfo
  {volume} {16}},\ \bibinfo {pages} {4192} (\bibinfo {year}
  {2020})}\BibitemShut {NoStop}%
\bibitem [{\citenamefont {Smith}\ \emph {et~al.}(2019)\citenamefont {Smith},
  \citenamefont {Nebgen}, \citenamefont {Zubatyuk}, \citenamefont {Lubbers},
  \citenamefont {Devereux}, \citenamefont {Barros}, \citenamefont {Tretiak},
  \citenamefont {Isayev},\ and\ \citenamefont {Roitberg}}]{ani2}%
  \BibitemOpen
  \bibfield  {author} {\bibinfo {author} {\bibfnamefont {J.~S.}\ \bibnamefont
  {Smith}}, \bibinfo {author} {\bibfnamefont {B.~T.}\ \bibnamefont {Nebgen}},
  \bibinfo {author} {\bibfnamefont {R.}~\bibnamefont {Zubatyuk}}, \bibinfo
  {author} {\bibfnamefont {N.}~\bibnamefont {Lubbers}}, \bibinfo {author}
  {\bibfnamefont {C.}~\bibnamefont {Devereux}}, \bibinfo {author}
  {\bibfnamefont {K.}~\bibnamefont {Barros}}, \bibinfo {author} {\bibfnamefont
  {S.}~\bibnamefont {Tretiak}}, \bibinfo {author} {\bibfnamefont
  {O.}~\bibnamefont {Isayev}}, \ and\ \bibinfo {author} {\bibfnamefont {A.~E.}\
  \bibnamefont {Roitberg}},\ }\href@noop {} {\bibfield  {journal} {\bibinfo
  {journal} {Nat. Commun}\ }\textbf {\bibinfo {volume} {10}} (\bibinfo {year}
  {2019})}\BibitemShut {NoStop}%
\bibitem [{\citenamefont {Smith}\ \emph {et~al.}(2018)\citenamefont {Smith},
  \citenamefont {Nebgen},\ and\ \citenamefont {Lubbers}}]{ani3}%
  \BibitemOpen
  \bibfield  {author} {\bibinfo {author} {\bibfnamefont {J.~S.}\ \bibnamefont
  {Smith}}, \bibinfo {author} {\bibfnamefont {B.}~\bibnamefont {Nebgen}}, \
  and\ \bibinfo {author} {\bibfnamefont {N.}~\bibnamefont {Lubbers}},\
  }\href@noop {} {\bibfield  {journal} {\bibinfo  {journal} {J. Chem. Phys.}\
  }\textbf {\bibinfo {volume} {148}} (\bibinfo {year} {2018})}\BibitemShut
  {NoStop}%
\bibitem [{\citenamefont {Glielmo}\ \emph {et~al.}(2017)\citenamefont
  {Glielmo}, \citenamefont {Sollich},\ and\ \citenamefont
  {De~Vita}}]{glielmo2017accurate}%
  \BibitemOpen
  \bibfield  {author} {\bibinfo {author} {\bibfnamefont {A.}~\bibnamefont
  {Glielmo}}, \bibinfo {author} {\bibfnamefont {P.}~\bibnamefont {Sollich}}, \
  and\ \bibinfo {author} {\bibfnamefont {A.}~\bibnamefont {De~Vita}},\
  }\href@noop {} {\bibfield  {journal} {\bibinfo  {journal} {Phys. Rev. B}\
  }\textbf {\bibinfo {volume} {95}},\ \bibinfo {pages} {214302} (\bibinfo
  {year} {2017})}\BibitemShut {NoStop}%
\bibitem [{\citenamefont {Grisafi}\ \emph {et~al.}(2018)\citenamefont
  {Grisafi}, \citenamefont {Wilkins}, \citenamefont {Cs\'{a}nyi},\ and\
  \citenamefont {Ceriotti}}]{ceriottis-sagpr}%
  \BibitemOpen
  \bibfield  {author} {\bibinfo {author} {\bibfnamefont {A.}~\bibnamefont
  {Grisafi}}, \bibinfo {author} {\bibfnamefont {D.~M.}\ \bibnamefont
  {Wilkins}}, \bibinfo {author} {\bibfnamefont {G.}~\bibnamefont {Cs\'{a}nyi}},
  \ and\ \bibinfo {author} {\bibfnamefont {M.}~\bibnamefont {Ceriotti}},\
  }\href@noop {} {\bibfield  {journal} {\bibinfo  {journal} {Phys. Rev. Lett.}\
  }\textbf {\bibinfo {volume} {120}} (\bibinfo {year} {2018})}\BibitemShut
  {NoStop}%
\bibitem [{\citenamefont {Lunghi}\ and\ \citenamefont
  {Sanvito}(2020{\natexlab{a}})}]{ales-jpcc}%
  \BibitemOpen
  \bibfield  {author} {\bibinfo {author} {\bibfnamefont {A.}~\bibnamefont
  {Lunghi}}\ and\ \bibinfo {author} {\bibfnamefont {S.}~\bibnamefont
  {Sanvito}},\ }\href@noop {} {\bibfield  {journal} {\bibinfo  {journal} {J.
  Phys. Chem. C}\ }\textbf {\bibinfo {volume} {124}},\ \bibinfo {pages}
  {5802−5806} (\bibinfo {year} {2020}{\natexlab{a}})}\BibitemShut {NoStop}%
\bibitem [{\citenamefont {Lunghi}(2020)}]{lunghi2020insights}%
  \BibitemOpen
  \bibfield  {author} {\bibinfo {author} {\bibfnamefont {A.}~\bibnamefont
  {Lunghi}},\ }\href@noop {} {\bibfield  {journal} {\bibinfo  {journal} {Appl.
  Mag. Reson.}\ }\textbf {\bibinfo {volume} {51}},\ \bibinfo {pages} {1343}
  (\bibinfo {year} {2020})}\BibitemShut {NoStop}%
\bibitem [{\citenamefont {Zaverkin}\ \emph {et~al.}(2021)\citenamefont
  {Zaverkin}, \citenamefont {Netz}, \citenamefont {Zills}, \citenamefont
  {K\"{o}hn},\ and\ \citenamefont {K\"{a}stner}}]{gauss-moment}%
  \BibitemOpen
  \bibfield  {author} {\bibinfo {author} {\bibfnamefont {V.}~\bibnamefont
  {Zaverkin}}, \bibinfo {author} {\bibfnamefont {J.}~\bibnamefont {Netz}},
  \bibinfo {author} {\bibfnamefont {F.}~\bibnamefont {Zills}}, \bibinfo
  {author} {\bibfnamefont {A.}~\bibnamefont {K\"{o}hn}}, \ and\ \bibinfo
  {author} {\bibfnamefont {J.}~\bibnamefont {K\"{a}stner}},\ }\href@noop {}
  {\bibfield  {journal} {\bibinfo  {journal} {J. Chem. Theory Comput.}\ }
  (\bibinfo {year} {2021})}\BibitemShut {NoStop}%
\bibitem [{\citenamefont {Thompson}\ \emph {et~al.}(2015)\citenamefont
  {Thompson}, \citenamefont {Swiler}, \citenamefont {Trott}, \citenamefont
  {Foiles},\ and\ \citenamefont {Tucker}}]{snap2}%
  \BibitemOpen
  \bibfield  {author} {\bibinfo {author} {\bibfnamefont {A.~P.}\ \bibnamefont
  {Thompson}}, \bibinfo {author} {\bibfnamefont {L.~P.}\ \bibnamefont
  {Swiler}}, \bibinfo {author} {\bibfnamefont {C.~R.}\ \bibnamefont {Trott}},
  \bibinfo {author} {\bibfnamefont {S.~M.}\ \bibnamefont {Foiles}}, \ and\
  \bibinfo {author} {\bibfnamefont {G.~J.}\ \bibnamefont {Tucker}},\
  }\href@noop {} {\bibfield  {journal} {\bibinfo  {journal} {J. Comput. Phys.}\
  }\textbf {\bibinfo {volume} {285}},\ \bibinfo {pages} {316} (\bibinfo {year}
  {2015})}\BibitemShut {NoStop}%
\bibitem [{\citenamefont {Lunghi}\ and\ \citenamefont {Sanvito}(2019)}]{snap1}%
  \BibitemOpen
  \bibfield  {author} {\bibinfo {author} {\bibfnamefont {A.}~\bibnamefont
  {Lunghi}}\ and\ \bibinfo {author} {\bibfnamefont {S.}~\bibnamefont
  {Sanvito}},\ }\href@noop {} {\bibfield  {journal} {\bibinfo  {journal} {Sci.
  Adv.}\ }\textbf {\bibinfo {volume} {5}} (\bibinfo {year} {2019})}\BibitemShut
  {NoStop}%
\bibitem [{\citenamefont {Bart\'{o}k}\ \emph {et~al.}(2013)\citenamefont
  {Bart\'{o}k}, \citenamefont {Kondor},\ and\ \citenamefont
  {Cs\'{a}nyi}}]{representation}%
  \BibitemOpen
  \bibfield  {author} {\bibinfo {author} {\bibfnamefont {A.~P.}\ \bibnamefont
  {Bart\'{o}k}}, \bibinfo {author} {\bibfnamefont {R.}~\bibnamefont {Kondor}},
  \ and\ \bibinfo {author} {\bibfnamefont {G.}~\bibnamefont {Cs\'{a}nyi}},\
  }\href@noop {} {\bibfield  {journal} {\bibinfo  {journal} {Phys. Rev. B}\
  }\textbf {\bibinfo {volume} {87}} (\bibinfo {year} {2013})}\BibitemShut
  {NoStop}%
\bibitem [{\citenamefont {Lunghi}\ and\ \citenamefont
  {Sanvito}(2020{\natexlab{b}})}]{lunghi2020limit}%
  \BibitemOpen
  \bibfield  {author} {\bibinfo {author} {\bibfnamefont {A.}~\bibnamefont
  {Lunghi}}\ and\ \bibinfo {author} {\bibfnamefont {S.}~\bibnamefont
  {Sanvito}},\ }\href@noop {} {\bibfield  {journal} {\bibinfo  {journal} {J.
  Phys. Chem. Lett.}\ }\textbf {\bibinfo {volume} {11}},\ \bibinfo {pages}
  {6273} (\bibinfo {year} {2020}{\natexlab{b}})}\BibitemShut {NoStop}%
\bibitem [{\citenamefont {Thomas}\ \emph {et~al.}(2018)\citenamefont {Thomas},
  \citenamefont {Smidt}, \citenamefont {Kearnes}, \citenamefont {Yang},
  \citenamefont {Li}, \citenamefont {Kohlhoff},\ and\ \citenamefont
  {Riley}}]{tensor-field}%
  \BibitemOpen
  \bibfield  {author} {\bibinfo {author} {\bibfnamefont {N.}~\bibnamefont
  {Thomas}}, \bibinfo {author} {\bibfnamefont {T.}~\bibnamefont {Smidt}},
  \bibinfo {author} {\bibfnamefont {S.}~\bibnamefont {Kearnes}}, \bibinfo
  {author} {\bibfnamefont {L.}~\bibnamefont {Yang}}, \bibinfo {author}
  {\bibfnamefont {L.}~\bibnamefont {Li}}, \bibinfo {author} {\bibfnamefont
  {K.}~\bibnamefont {Kohlhoff}}, \ and\ \bibinfo {author} {\bibfnamefont
  {P.}~\bibnamefont {Riley}},\ }\href@noop {} {\enquote {\bibinfo {title}
  {Tensor field networks: Rotation- and translation-equivariant neural networks
  for 3d point clouds},}\ } (\bibinfo {year} {2018}),\ \Eprint
  {http://arxiv.org/abs/1802.08219} {arXiv:1802.08219} \BibitemShut {NoStop}%
\bibitem [{\citenamefont {Weiler}\ \emph {et~al.}(2018)\citenamefont {Weiler},
  \citenamefont {Geiger}, \citenamefont {Welling}, \citenamefont {Boomsma},\
  and\ \citenamefont {Cohen}}]{3d-steerable}%
  \BibitemOpen
  \bibfield  {author} {\bibinfo {author} {\bibfnamefont {M.}~\bibnamefont
  {Weiler}}, \bibinfo {author} {\bibfnamefont {M.}~\bibnamefont {Geiger}},
  \bibinfo {author} {\bibfnamefont {M.}~\bibnamefont {Welling}}, \bibinfo
  {author} {\bibfnamefont {W.}~\bibnamefont {Boomsma}}, \ and\ \bibinfo
  {author} {\bibfnamefont {T.}~\bibnamefont {Cohen}},\ }\href@noop {} {\enquote
  {\bibinfo {title} {3d steerable cnns: Learning rotationally equivariant
  features in volumetric data},}\ } (\bibinfo {year} {2018}),\ \Eprint
  {http://arxiv.org/abs/1807.02547} {arXiv:1807.02547} \BibitemShut {NoStop}%
\bibitem [{\citenamefont {Kondor}\ \emph {et~al.}(2018)\citenamefont {Kondor},
  \citenamefont {Lin},\ and\ \citenamefont {Trivedi}}]{clebsch-gordan-nets}%
  \BibitemOpen
  \bibfield  {author} {\bibinfo {author} {\bibfnamefont {R.}~\bibnamefont
  {Kondor}}, \bibinfo {author} {\bibfnamefont {Z.}~\bibnamefont {Lin}}, \ and\
  \bibinfo {author} {\bibfnamefont {S.}~\bibnamefont {Trivedi}},\ }in\ \href
  {http://papers.nips.cc/paper/8215-clebschgordan-nets-a-fully-fourier-space-spherical-convolutional-neural-network}
  {\emph {\bibinfo {booktitle} {NeurIPS}}}\ (\bibinfo {year} {2018})\ pp.\
  \bibinfo {pages} {10138--10147}\BibitemShut {NoStop}%
\bibitem [{\citenamefont {Anderson}\ \emph {et~al.}(2019)\citenamefont
  {Anderson}, \citenamefont {Hy},\ and\ \citenamefont {Kondor}}]{cormorant}%
  \BibitemOpen
  \bibfield  {author} {\bibinfo {author} {\bibfnamefont {B.}~\bibnamefont
  {Anderson}}, \bibinfo {author} {\bibfnamefont {T.-S.}\ \bibnamefont {Hy}}, \
  and\ \bibinfo {author} {\bibfnamefont {R.}~\bibnamefont {Kondor}},\
  }\href@noop {} {\enquote {\bibinfo {title} {Cormorant: Covariant molecular
  neural networks},}\ } (\bibinfo {year} {2019}),\ \Eprint
  {http://arxiv.org/abs/1906.04015} {arXiv:1906.04015} \BibitemShut {NoStop}%
\bibitem [{\citenamefont {Cohen}\ and\ \citenamefont
  {Welling}(2016)}]{group-equiv-conv}%
  \BibitemOpen
  \bibfield  {author} {\bibinfo {author} {\bibfnamefont {T.}~\bibnamefont
  {Cohen}}\ and\ \bibinfo {author} {\bibfnamefont {M.}~\bibnamefont
  {Welling}},\ }in\ \href@noop {} {\emph {\bibinfo {booktitle} {Proceedings of
  The 33rd International Conference on Machine Learning}}},\ \bibinfo {series}
  {Proceedings of Machine Learning Research}, Vol.~\bibinfo {volume} {48},\
  \bibinfo {editor} {edited by\ \bibinfo {editor} {\bibfnamefont {M.~F.}\
  \bibnamefont {Balcan}}\ and\ \bibinfo {editor} {\bibfnamefont {K.~Q.}\
  \bibnamefont {Weinberger}}}\ (\bibinfo  {publisher} {PMLR},\ \bibinfo
  {address} {New York, New York, USA},\ \bibinfo {year} {2016})\ pp.\ \bibinfo
  {pages} {2990--2999}\BibitemShut {NoStop}%
\bibitem [{\citenamefont {Shapeev}(2016)}]{moment-tensor-potentials}%
  \BibitemOpen
  \bibfield  {author} {\bibinfo {author} {\bibfnamefont {A.~V.}\ \bibnamefont
  {Shapeev}},\ }\href@noop {} {\bibfield  {journal} {\bibinfo  {journal}
  {Multiscale Model. Simul.}\ }\textbf {\bibinfo {volume} {14}},\ \bibinfo
  {pages} {1153–1173} (\bibinfo {year} {2016})}\BibitemShut {NoStop}%
\bibitem [{\citenamefont {Langer}\ \emph {et~al.}(2021)\citenamefont {Langer},
  \citenamefont {Goe{\ss}mann},\ and\ \citenamefont
  {Rupp}}]{langer2021representations}%
  \BibitemOpen
  \bibfield  {author} {\bibinfo {author} {\bibfnamefont {M.~F.}\ \bibnamefont
  {Langer}}, \bibinfo {author} {\bibfnamefont {A.}~\bibnamefont
  {Goe{\ss}mann}}, \ and\ \bibinfo {author} {\bibfnamefont {M.}~\bibnamefont
  {Rupp}},\ }\href@noop {} {\enquote {\bibinfo {title} {Representations of
  molecules and materials for interpolation of quantum-mechanical simulations
  via machine learning},}\ } (\bibinfo {year} {2021}),\ \Eprint
  {http://arxiv.org/abs/2003.12081} {arXiv:2003.12081 [physics.comp-ph]}
  \BibitemShut {NoStop}%
\bibitem [{\citenamefont {Miller}\ \emph {et~al.}(2020)\citenamefont {Miller},
  \citenamefont {Geiger}, \citenamefont {Smidt},\ and\ \citenamefont
  {No\'{e}}}]{enn}%
  \BibitemOpen
  \bibfield  {author} {\bibinfo {author} {\bibfnamefont {B.~K.}\ \bibnamefont
  {Miller}}, \bibinfo {author} {\bibfnamefont {M.}~\bibnamefont {Geiger}},
  \bibinfo {author} {\bibfnamefont {T.~E.}\ \bibnamefont {Smidt}}, \ and\
  \bibinfo {author} {\bibfnamefont {F.}~\bibnamefont {No\'{e}}},\ }\href@noop
  {} {\enquote {\bibinfo {title} {Relevance of rotationally equivariant
  convolutions for predicting molecular properties},}\ } (\bibinfo {year}
  {2020}),\ \Eprint {http://arxiv.org/abs/2008.08461} {arXiv:2008.08461
  [cs.LG]} \BibitemShut {NoStop}%
\bibitem [{\citenamefont {Zhang}\ \emph {et~al.}(2021)\citenamefont {Zhang},
  \citenamefont {Onat}, \citenamefont {Dusson}, \citenamefont {Anand},
  \citenamefont {Maurer}, \citenamefont {Ortner},\ and\ \citenamefont
  {Kermode}}]{zhang2021equivariant}%
  \BibitemOpen
  \bibfield  {author} {\bibinfo {author} {\bibfnamefont {L.}~\bibnamefont
  {Zhang}}, \bibinfo {author} {\bibfnamefont {B.}~\bibnamefont {Onat}},
  \bibinfo {author} {\bibfnamefont {G.}~\bibnamefont {Dusson}}, \bibinfo
  {author} {\bibfnamefont {G.}~\bibnamefont {Anand}}, \bibinfo {author}
  {\bibfnamefont {R.~J.}\ \bibnamefont {Maurer}}, \bibinfo {author}
  {\bibfnamefont {C.}~\bibnamefont {Ortner}}, \ and\ \bibinfo {author}
  {\bibfnamefont {J.~R.}\ \bibnamefont {Kermode}},\ }\href@noop {} {\bibfield
  {journal} {\bibinfo  {journal} {arXiv preprint arXiv:2111.13736}\ } (\bibinfo
  {year} {2021})}\BibitemShut {NoStop}%
\bibitem [{\citenamefont {Unke}\ \emph {et~al.}(2021)\citenamefont {Unke},
  \citenamefont {Bogojeski}, \citenamefont {Gastegger}, \citenamefont {Geiger},
  \citenamefont {Smidt},\ and\ \citenamefont {M{\"u}ller}}]{unke2021se}%
  \BibitemOpen
  \bibfield  {author} {\bibinfo {author} {\bibfnamefont {O.}~\bibnamefont
  {Unke}}, \bibinfo {author} {\bibfnamefont {M.}~\bibnamefont {Bogojeski}},
  \bibinfo {author} {\bibfnamefont {M.}~\bibnamefont {Gastegger}}, \bibinfo
  {author} {\bibfnamefont {M.}~\bibnamefont {Geiger}}, \bibinfo {author}
  {\bibfnamefont {T.}~\bibnamefont {Smidt}}, \ and\ \bibinfo {author}
  {\bibfnamefont {K.-R.}\ \bibnamefont {M{\"u}ller}},\ }\href@noop {}
  {\bibfield  {journal} {\bibinfo  {journal} {Advances in Neural Information
  Processing Systems}\ }\textbf {\bibinfo {volume} {34}} (\bibinfo {year}
  {2021})}\BibitemShut {NoStop}%
\bibitem [{\citenamefont {Nigam}\ \emph {et~al.}(2022)\citenamefont {Nigam},
  \citenamefont {Willatt},\ and\ \citenamefont
  {Ceriotti}}]{nigam2022equivariant}%
  \BibitemOpen
  \bibfield  {author} {\bibinfo {author} {\bibfnamefont {J.}~\bibnamefont
  {Nigam}}, \bibinfo {author} {\bibfnamefont {M.~J.}\ \bibnamefont {Willatt}},
  \ and\ \bibinfo {author} {\bibfnamefont {M.}~\bibnamefont {Ceriotti}},\
  }\href@noop {} {\bibfield  {journal} {\bibinfo  {journal} {J. Chem. Phys.}\
  }\textbf {\bibinfo {volume} {156}},\ \bibinfo {pages} {014115} (\bibinfo
  {year} {2022})}\BibitemShut {NoStop}%
\bibitem [{\citenamefont {Najafabadi}\ \emph {et~al.}(2015)\citenamefont
  {Najafabadi}, \citenamefont {Villanustre}, \citenamefont {Khoshgoftaar},
  \citenamefont {Seliya}, \citenamefont {Wald},\ and\ \citenamefont
  {Muharemagic}}]{nn-large-data}%
  \BibitemOpen
  \bibfield  {author} {\bibinfo {author} {\bibfnamefont {M.~M.}\ \bibnamefont
  {Najafabadi}}, \bibinfo {author} {\bibfnamefont {F.}~\bibnamefont
  {Villanustre}}, \bibinfo {author} {\bibfnamefont {T.~M.}\ \bibnamefont
  {Khoshgoftaar}}, \bibinfo {author} {\bibfnamefont {N.}~\bibnamefont
  {Seliya}}, \bibinfo {author} {\bibfnamefont {R.}~\bibnamefont {Wald}}, \ and\
  \bibinfo {author} {\bibfnamefont {E.}~\bibnamefont {Muharemagic}},\
  }\href@noop {} {\bibfield  {journal} {\bibinfo  {journal} {J. Big Data}\
  }\textbf {\bibinfo {volume} {2}} (\bibinfo {year} {2015})}\BibitemShut
  {NoStop}%
\bibitem [{\citenamefont {Perakis}\ \emph {et~al.}(2016)\citenamefont
  {Perakis}, \citenamefont {Marco}, \citenamefont {Shalit}, \citenamefont
  {Tang}, \citenamefont {Kann}, \citenamefont {K\"{u}hne}, \citenamefont
  {Torre}, \citenamefont {Bonn},\ and\ \citenamefont {Nagata}}]{ir-raman}%
  \BibitemOpen
  \bibfield  {author} {\bibinfo {author} {\bibfnamefont {F.}~\bibnamefont
  {Perakis}}, \bibinfo {author} {\bibfnamefont {L.~D.}\ \bibnamefont {Marco}},
  \bibinfo {author} {\bibfnamefont {A.}~\bibnamefont {Shalit}}, \bibinfo
  {author} {\bibfnamefont {F.}~\bibnamefont {Tang}}, \bibinfo {author}
  {\bibfnamefont {Z.~R.}\ \bibnamefont {Kann}}, \bibinfo {author}
  {\bibfnamefont {T.~D.}\ \bibnamefont {K\"{u}hne}}, \bibinfo {author}
  {\bibfnamefont {R.}~\bibnamefont {Torre}}, \bibinfo {author} {\bibfnamefont
  {M.}~\bibnamefont {Bonn}}, \ and\ \bibinfo {author} {\bibfnamefont
  {Y.}~\bibnamefont {Nagata}},\ }\href@noop {} {\bibfield  {journal} {\bibinfo
  {journal} {Chem. Rev.}\ }\textbf {\bibinfo {volume} {116}},\ \bibinfo {pages}
  {7590–7607} (\bibinfo {year} {2016})}\BibitemShut {NoStop}%
\bibitem [{\citenamefont {Lubera}\ \emph {et~al.}(2014)\citenamefont {Lubera},
  \citenamefont {Iannuzzi},\ and\ \citenamefont {Hutter}}]{ir-raman2}%
  \BibitemOpen
  \bibfield  {author} {\bibinfo {author} {\bibfnamefont {S.}~\bibnamefont
  {Lubera}}, \bibinfo {author} {\bibfnamefont {M.}~\bibnamefont {Iannuzzi}}, \
  and\ \bibinfo {author} {\bibfnamefont {J.}~\bibnamefont {Hutter}},\
  }\href@noop {} {\bibfield  {journal} {\bibinfo  {journal} {J. Chem. Phys.}\
  }\textbf {\bibinfo {volume} {141}},\ \bibinfo {pages} {094503} (\bibinfo
  {year} {2014})}\BibitemShut {NoStop}%
\bibitem [{\citenamefont {Briganti}\ \emph {et~al.}(2021)\citenamefont
  {Briganti}, \citenamefont {Santanni}, \citenamefont {Tesi}, \citenamefont
  {Totti}, \citenamefont {Sessoli},\ and\ \citenamefont
  {Lunghi}}]{briganti2021}%
  \BibitemOpen
  \bibfield  {author} {\bibinfo {author} {\bibfnamefont {M.}~\bibnamefont
  {Briganti}}, \bibinfo {author} {\bibfnamefont {F.}~\bibnamefont {Santanni}},
  \bibinfo {author} {\bibfnamefont {L.}~\bibnamefont {Tesi}}, \bibinfo {author}
  {\bibfnamefont {F.}~\bibnamefont {Totti}}, \bibinfo {author} {\bibfnamefont
  {R.}~\bibnamefont {Sessoli}}, \ and\ \bibinfo {author} {\bibfnamefont
  {A.}~\bibnamefont {Lunghi}},\ }\href@noop {} {\bibfield  {journal} {\bibinfo
  {journal} {J. Am. Chem. Soc.}\ }\textbf {\bibinfo {volume} {143}},\ \bibinfo
  {pages} {13633–13645} (\bibinfo {year} {2021})}\BibitemShut {NoStop}%
\bibitem [{\citenamefont {Escalera-Moreno}\ \emph {et~al.}(2018)\citenamefont
  {Escalera-Moreno}, \citenamefont {Baldov{\'i}}, \citenamefont {{n}o},\ and\
  \citenamefont {Coronado}}]{magnetic-appl2}%
  \BibitemOpen
  \bibfield  {author} {\bibinfo {author} {\bibfnamefont {L.}~\bibnamefont
  {Escalera-Moreno}}, \bibinfo {author} {\bibfnamefont {J.~J.}\ \bibnamefont
  {Baldov{\'i}}}, \bibinfo {author} {\bibfnamefont {A.~G.-A.}\ \bibnamefont
  {{n}o}}, \ and\ \bibinfo {author} {\bibfnamefont {E.}~\bibnamefont
  {Coronado}},\ }\href@noop {} {\bibfield  {journal} {\bibinfo  {journal}
  {Chem. Sci.}\ }\textbf {\bibinfo {volume} {9}},\ \bibinfo {pages} {3265}
  (\bibinfo {year} {2018})}\BibitemShut {NoStop}%
\bibitem [{\citenamefont {Neese}\ \emph {et~al.}(2019)\citenamefont {Neese},
  \citenamefont {Atanasov}, \citenamefont {Bistoni}, \citenamefont {Maganas},\
  and\ \citenamefont {Ye}}]{neese2019chemistry}%
  \BibitemOpen
  \bibfield  {author} {\bibinfo {author} {\bibfnamefont {F.}~\bibnamefont
  {Neese}}, \bibinfo {author} {\bibfnamefont {M.}~\bibnamefont {Atanasov}},
  \bibinfo {author} {\bibfnamefont {G.}~\bibnamefont {Bistoni}}, \bibinfo
  {author} {\bibfnamefont {D.}~\bibnamefont {Maganas}}, \ and\ \bibinfo
  {author} {\bibfnamefont {S.}~\bibnamefont {Ye}},\ }\href@noop {} {\bibfield
  {journal} {\bibinfo  {journal} {J. Am. Chem. Soc.}\ }\textbf {\bibinfo
  {volume} {141}},\ \bibinfo {pages} {2814} (\bibinfo {year}
  {2019})}\BibitemShut {NoStop}%
\bibitem [{\citenamefont {Rechkemmer}\ \emph {et~al.}(2016)\citenamefont
  {Rechkemmer}, \citenamefont {Breitgoff}, \citenamefont {van~der Meer},
  \citenamefont {Atanasov}, \citenamefont {Hakl}, \citenamefont {Orlita},
  \citenamefont {Neugebauer}, \citenamefont {Neese}, \citenamefont {Sarkar},\
  and\ \citenamefont {van Slageren}}]{copdms}%
  \BibitemOpen
  \bibfield  {author} {\bibinfo {author} {\bibfnamefont {Y.}~\bibnamefont
  {Rechkemmer}}, \bibinfo {author} {\bibfnamefont {F.~D.}\ \bibnamefont
  {Breitgoff}}, \bibinfo {author} {\bibfnamefont {M.}~\bibnamefont {van~der
  Meer}}, \bibinfo {author} {\bibfnamefont {M.}~\bibnamefont {Atanasov}},
  \bibinfo {author} {\bibfnamefont {M.}~\bibnamefont {Hakl}}, \bibinfo {author}
  {\bibfnamefont {M.}~\bibnamefont {Orlita}}, \bibinfo {author} {\bibfnamefont
  {P.}~\bibnamefont {Neugebauer}}, \bibinfo {author} {\bibfnamefont
  {F.}~\bibnamefont {Neese}}, \bibinfo {author} {\bibfnamefont
  {B.}~\bibnamefont {Sarkar}}, \ and\ \bibinfo {author} {\bibfnamefont
  {J.}~\bibnamefont {van Slageren}},\ }\href@noop {} {\bibfield  {journal}
  {\bibinfo  {journal} {Nat. Commun}\ }\textbf {\bibinfo {volume} {7}}
  (\bibinfo {year} {2016})}\BibitemShut {NoStop}%
\bibitem [{\citenamefont {Lunghi}\ and\ \citenamefont
  {Sanvito}(2020{\natexlab{c}})}]{lunghi2020multiple}%
  \BibitemOpen
  \bibfield  {author} {\bibinfo {author} {\bibfnamefont {A.}~\bibnamefont
  {Lunghi}}\ and\ \bibinfo {author} {\bibfnamefont {S.}~\bibnamefont
  {Sanvito}},\ }\href@noop {} {\bibfield  {journal} {\bibinfo  {journal}
  {Journal Chem. Phys.}\ }\textbf {\bibinfo {volume} {153}},\ \bibinfo {pages}
  {174113} (\bibinfo {year} {2020}{\natexlab{c}})}\BibitemShut {NoStop}%
\bibitem [{\citenamefont {Chibotaru}\ and\ \citenamefont
  {Ungur}(2012)}]{ungur-cf}%
  \BibitemOpen
  \bibfield  {author} {\bibinfo {author} {\bibfnamefont {L.~F.}\ \bibnamefont
  {Chibotaru}}\ and\ \bibinfo {author} {\bibfnamefont {L.}~\bibnamefont
  {Ungur}},\ }\href@noop {} {\bibfield  {journal} {\bibinfo  {journal} {J.
  Chem. Phys.}\ }\textbf {\bibinfo {volume} {137}},\ \bibinfo {pages} {064112}
  (\bibinfo {year} {2012})}\BibitemShut {NoStop}%
\bibitem [{\citenamefont {Ryabov}(1999)}]{ryabov-1999}%
  \BibitemOpen
  \bibfield  {author} {\bibinfo {author} {\bibfnamefont {I.~D.}\ \bibnamefont
  {Ryabov}},\ }\href@noop {} {\bibfield  {journal} {\bibinfo  {journal} {J.
  Magn. Reson.}\ }\textbf {\bibinfo {volume} {140}},\ \bibinfo {pages} {141}
  (\bibinfo {year} {1999})}\BibitemShut {NoStop}%
\bibitem [{\citenamefont {Court}\ and\ \citenamefont
  {Cole}(2020)}]{ml-mag-bulk}%
  \BibitemOpen
  \bibfield  {author} {\bibinfo {author} {\bibfnamefont {C.~J.}\ \bibnamefont
  {Court}}\ and\ \bibinfo {author} {\bibfnamefont {J.~M.}\ \bibnamefont
  {Cole}},\ }\href@noop {} {\bibfield  {journal} {\bibinfo  {journal} {Npj
  Comput. Mater.}\ }\textbf {\bibinfo {volume} {6}} (\bibinfo {year}
  {2020})}\BibitemShut {NoStop}%
\bibitem [{\citenamefont {Paruzzo}\ \emph
  {et~al.}(2018{\natexlab{b}})\citenamefont {Paruzzo}, \citenamefont
  {Hofstetter}, \citenamefont {Musil}, \citenamefont {De}, \citenamefont
  {Ceriotti},\ and\ \citenamefont {Emsley}}]{paruzzo2018chemical}%
  \BibitemOpen
  \bibfield  {author} {\bibinfo {author} {\bibfnamefont {F.~M.}\ \bibnamefont
  {Paruzzo}}, \bibinfo {author} {\bibfnamefont {A.}~\bibnamefont {Hofstetter}},
  \bibinfo {author} {\bibfnamefont {F.}~\bibnamefont {Musil}}, \bibinfo
  {author} {\bibfnamefont {S.}~\bibnamefont {De}}, \bibinfo {author}
  {\bibfnamefont {M.}~\bibnamefont {Ceriotti}}, \ and\ \bibinfo {author}
  {\bibfnamefont {L.}~\bibnamefont {Emsley}},\ }\href@noop {} {\bibfield
  {journal} {\bibinfo  {journal} {Nat. Commun.}\ }\textbf {\bibinfo {volume}
  {9}},\ \bibinfo {pages} {4501} (\bibinfo {year}
  {2018}{\natexlab{b}})}\BibitemShut {NoStop}%
\end{thebibliography}

\end{document}